\documentclass[aps,prl,preprint,groupedaddress]{revtex4}

\usepackage{graphicx}
\usepackage{amsbsy,amssymb}
\usepackage{amsmath}
\usepackage{epsf}
\usepackage[usenames,dvips]{color}
\newcommand{\be}{\begin{equation}}
\newcommand{\ee}{\end{equation}}
\newcommand{\bea}{\begin{eqnarray}}
\newcommand{\eea}{\end{eqnarray}}
\newcommand{\LTO}{LiTi$_2$O$_4$}
\newcommand{\LiTiO}{LiTi$_2$O$_4$}
\newcommand{\LiVO}{LiV$_2$O$_4$}

\newcommand{\LiAlTiO}{LiAl$_y$Ti$_{2-y}$O$_4$}
                                                                                                                                           
\newcommand{\is}{i \sigma}
\newcommand{\js}{j \sigma}
\newcommand{\iu}{i\uparrow}
\newcommand{\id}{i\downarrow}
\newcommand{\as}{\alpha \sigma}
\newcommand{\eps}{\varepsilon}
\newcommand{\lgl}{\langle}
\newcommand{\rgl}{\rangle}

\begin{document}

%\preprint{QCMP Theory, 03-06-1}

\title{The role of strong electronic correlations\\
        in the metal-to-insulator transition in disordered \LiAlTiO}



\author{F. Fazileh}
\author{R. J. Gooding}
\affiliation{Department of Physics, Queen's University,
                Kingston ON K7L 3N6 Canada}

\author{W. A. Atkinson}
\affiliation{Department of Physics, Trent University,
                Peterborough ON K9J 7B8 Canada}

\author{D. C. Johnston}
\affiliation{Ames Laboratory and Department of Physics and Astronomy, Iowa State University,
                Ames IA 50011}

%\thanks{This work was partially supported by NSERC of Canada and NATO.}


\date{\today}

\begin{abstract}
The compound LiAl$_y$Ti$_{2-y}$O$_4$ undergoes a metal-to-insulator transition
for $y_c\sim 0.33$. It is known that disorder alone is insufficient to explain
this transition; e.g., a quantum site percolation model predicts $y_c\sim 0.8$.
We have included (Hubbard) electronic interactions into a model of this compound,
using a real-space Hartree-Fock approach that acheives self consistency at
every site, and have found that for a Hubbard energy equal to 1.5 times the
non-interacting bandwidth one obtains $y_c\sim 0.3$. Further, with increasing
Hubbard energy we find an Altshuler-Aronov suppression of the density of states,
$\delta N(\epsilon)~\sim~\sqrt{|\epsilon-\epsilon_F|}$, that reduces the density
of states at the Fermi energy to zero at the critical Hubbard interaction.
Using this ratio of correlation to hopping energy one is led to a
prediction of the near-neighbour superexchange ($J/t\sim 1/3$) which is similar
to that for the cuprate superconductors.
\end{abstract}

\maketitle

The range of interesting phenomena of transition metal systems includes the anomalous normal state
and high-temperature superconductivity of the quasi-2d cuprates \cite{Bednorz86},
as well as the colossal magnetoresistance of the manganites and related systems \cite{CMR-Dagotto}.
The ubiquitous physics believed to be responsible for these novel behaviours is strong
electron-electron interactions, a consequence of which is the inapplicability of Landau's
theory of fermi liquids.
One interesting class of related materials are the spinels. There are over 300 transition
metal spinels, out of which only four are superconducting. Further, of these four only one
is an oxide, and that oxide, \LiTiO, has the highest superconducting transition temperature,
$T_c$ \cite{Moshopoulou99}: CuRh$_2$Se$_4$~(T$_c = 3.49$~K), CuV$_2$S$_4$~(T$_c = 4.45$~K),
CuRh$_2$S$_4$~(T$_c = 4.8$~K), and \LiTiO~(T$_c = 11.3~K$).

\LiTiO~ is a 1/4-filled $d^{0.5}$ system in which the electronic conduction 
occurs via direct $d-d$ hopping on the Ti sites, owing to the orientation of the low-lying
$t_{2g}$ orbitals\cite{Satpathy87,Massidda88}. The Ti sublattice corresponds to a corner-sharing 
tetrahedral lattice (CSTL), which is a fully frustrated three dimensional (3d) structure.
It has been suggested by Bednorz and M\"uller \cite{Muller96}
that this system is moderately correlated electronic system, and that the superconductivity
may be driven by the electronic interactions amongst the $d$ electrons. However, other
reasons behind our interest in this system include: (i) A metal-to-insulator transition generated
by both disorder and a reduced electronic density caused by chemically substituting Li, Al or Cr 
for Ti \cite{Johnston76I,Lambert90} -- this transition is the focus of this paper. 
(ii) If strong electronic correlations are present then the magnetic properties of this
material could be interesting, corresponding to a 1/4-filled fully frustrated lattice.
(iii) The isostructural $d^{1.5}$ \LiVO~
is believed to be a $d$-electron heavy fermion compound \cite{Kondo97}, and understanding
the simpler (1/2 an itinerant electron per site in \LiTiO~\textit{vs.} the
1.5 electron per site (1 local moment plus 1/2 an itinerant electron per site))
\LiVO~would be highly beneficial.

In the tight-binding approximation, a reasonable Hamiltonian from which
to begin a study of disordered \LiAlTiO~is given by
\begin{equation}
\label{eq:H_full}
\mathcal{H} = \sum_{i,\sigma} \varepsilon_i n_{i,\sigma}
-t~{\sum_{\lgl ij\rgl ,\sigma}}
(c_{\is}^\dagger c_{\js} + h.c.)~+~ U~{\sum_i} n_{\iu}n_{\id}
\end{equation}
where $i$ denotes a Ti site on a CSTL, and $c_{\is}$ the
annihilation operator for an electron at site $i$ with spin $\sigma$, and 
$n_{i,\sigma}~=~c_{\is}^\dagger c_{\is}$. (For reference below, note that the noninteracting
bandwidth is 8$t$.)

As mentioned above, there are a variety of different chemical dopings that lead to a 
metal-to-insulator transition (MIT), but here we focus on \LiAlTiO, which undergoes 
a MIT for $y\sim 0.33$ \cite{Lambert90}. In this system the Al$^{3+}$ ions substitute 
onto the Ti sublattice, and thus in a first approximation the on-site energies are chosen 
at random according to the distribution function
\begin{equation}
\label{eq:SiteDistr}
P(\varepsilon_i) = (1-\frac{y}{2})~\delta(\varepsilon_i-\varepsilon_{\rm Ti})+
\frac{y}{2}~\delta(\varepsilon_i-\varepsilon_{\rm Al}) \qquad .
\end{equation}
Then, in the expected limit of $\varepsilon_{\rm Al} - \varepsilon_{\rm Ti} \gg 8t$, 
\textit{if} one ignores the presence of electron-electron interactions one sees that 
this system is an excellent representation of quantum site percolation (QSP).
Previously, three of us \cite{Fazileh04} have analyzed such a QSP model for noninteracting 
electrons ($U=0$) on a CSTL, and (numerically) exactly solved the 
disordered electron problem (energies and eigenfunctions) for various large lattice sizes, 
thus determining the Fermi energy, $E_f(y)$, and mobility edge, $E_c(y)$, as a function of
Al doping concentrations. We found that at $y=y_c\sim 0.8$ the Fermi energy and mobility
edges crosses, and in such a disordered, noninteracting model this would correspond to the 
predicted MIT. The large disagreement between the experimental value of $y_c \sim 0.33$ and our
prediction ($y_c\sim 0.8$) highlights that important physics has been omitted in such an analysis
\cite{ACSTLcomment}.

Indirectly, the above result supports the conjecture that strong electronic correlations
are present and play an important role in the physics of \LiTiO. Of course, to provide
substance to this idea we need to examine the full disordered and interacting electron
problem, and to this end we have analyzed the behaviour of such a disordered-Hubbard
Hamiltonian, \textit{viz.} we have studied Eq.~(\ref{eq:H_full}) with the Ti conduction
path limited by the full QSP model mentioned above, now including the Hubbard interaction
term. This is a relatively complicated model which contains the interplay between electronic 
correlations and disorder produced by quantum-site-percolation, and in the following
we describe our results of a comprehensive examination of this problem in a real-space
self-consistent Hartree-Fock approximation. Recent results\cite{Trivedi04} for a 
disordered Hubbard model on a two-dimensional square lattice have lead to the 
interesting prediction of a novel metallic phase; here we use the same formalism, 
but now for a 3d CSTL.

In such a real-space self-consistent Hartree-Fock formulation one replaces the 
Hubbard interaction term as follows:
\begin{equation}
\label{eq:decompose3}
U{\sum_i}^{'} n_{\iu}n_{\id} \simeq U{\sum_i}^{'}(\lgl n_{\iu} \rgl n_{\id} 
+\lgl n_{\id} \rgl n_{\iu}-\lgl n_{\iu} \rgl \lgl n_{\id} \rgl)\quad ,
\end{equation}
where the primed summation indicates that only Ti sites (from both the maximally connected, 
and possibly isolated clusters) are included. Then the interacting Hamiltonian is simplified to
\begin{eqnarray}
\label{eq:SMFH1}
\mathcal{H}_{eff} = &-t&{\sum_{\lgl ij\rgl , \sigma}}^{'} c_{\is}^\dagger c_{\js}  
-U {\sum_i}^{'}\lgl n_{\iu} \rgl \lgl n_{\id} \rgl \\ &+& U {\sum_i}^{'}[\lgl n_{\iu}\rgl c_{\id}^\dagger 
c_{\id} + \lgl n_{\id}\rgl c_{\iu}^\dagger c_{\iu}] \nonumber
\end{eqnarray}
where
\begin{equation}
\label{eq:ni}
\lgl n_{\is}\rgl = \sum_{\alpha \atop \eps_{\as}\le \eps_F} |\lgl i|\psi_{\as}\rgl|^2 ~,~
\end{equation}
and
\begin{equation}
\mathcal{H}_{eff}|\psi_{\as}\rgl = \eps_{\as}|\psi_{\as}\rgl
\end{equation}
Here, $|i\rgl$ is a single orbital Wannier function at site $i$ from the Ti sublattice, 
$|\psi_{\as}\rgl$ are the energy eigenstates of the effective Hamiltonian, $\lgl n_{\is}\rgl$ is the 
average density of electrons with spin $\sigma$ at site $i$, and $\eps_F$ is the 
Fermi energy \cite{SPINcomment}.

In order to find self-consistent solutions for the above Hamiltonian, we start with the noninteracting
but disordered tight-binding model on a CSTL, and obtain the 
expectation values of the density of electrons on each site using Eq.~(\ref{eq:ni}) -- this is the distribution of the 
density of electrons for a non-interacting QSP disordered system for a specific filling factor of the \LiAlTiO~ system.  
We then iterate to convergence {\em at every site} as $U$ is increased from zero.
Note that in the iteration sequence we have added a small random fluctuation in the on-site densities for
each site and each spin, to allow for the system to proceed to a non-paramagnetic state if it so chooses.

This procedure has been applied to CSTL, with random QSP determined by $y$,
for system sizes of 5488, 8192, 11664 and 16000 lattice sites with periodic boundary conditions. 
For each system size the spectrum of eigenvalues and the distribution of the electron density throughout the lattice
were calculated for each realization of disorder, for doping concentrations of 
$y=0.2, 0.3, 0.4, 0.6, 0.7,$ and $0.75$, and for Hubbard interaction strengths of 
$U/t=2, 4, 6, 8, 10,$ and $12$. Then, these calculations were repeated for different
complexions of disorder for each set of values of $y$ and $U/t$.

\begin{figure}[t]
\begin{center}
\raisebox{-1cm}[6.5cm][1.5cm]{
\includegraphics[width=9.75cm,height=8.cm]{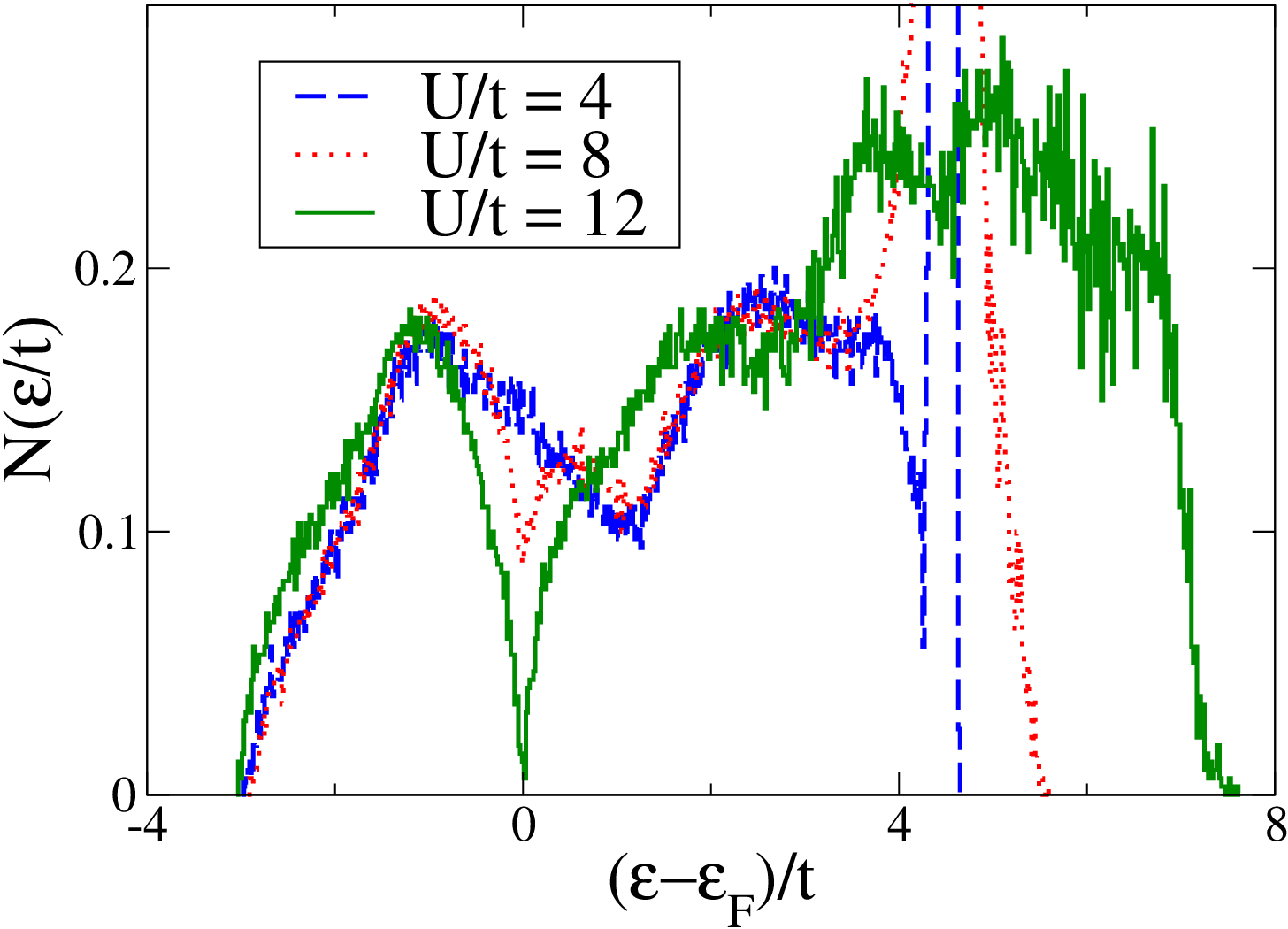}}
\caption{\label{fig:DOS_y0pt3}Density of states for a single realization of disorder in the Hubbard+QSP model for the 
\LiAlTiO~system with $y=0.3$ and $U/t=4,8,12$ for a system with 16000 lattice sites. The suppression that
is associated with the Altshuler-Aronov square-root singularity is evident, as is the complete suppression (zero density
of states at the Fermi level) at the critical $(U_c/t,y_c)$.}
\end{center}
\end{figure}
Density of states results are shown in Fig.~\ref{fig:DOS_y0pt3} for a doping concentration of $y=0.3$ and
for three different Hubbard interactions of $U/t=4,8,12$, for systems with 16000 lattice sites. 
The interesting feature in these plots is the suppression of the density of states at the Fermi level with 
increasing $U$.  That is, as a function of increasing Hubbard energy, we find that a progressively larger 
change of the density of states consistent with the predictions of the theory of Altshuler and Aronov 
\cite{Altshuler79}, namely a square-root suppression near the Fermi level of
the form $\delta N(\epsilon)~\sim~\sqrt{|\epsilon-\epsilon_F|}$. In fact, we find a complete suppression that 
\textit{first appears at the critical Hubbard interaction} (see below) associated with the
metal-to-insulator transition. These and our previous density of states data\cite{Fazileh04}
are consistent with the hypothesis that only in an interacting system should one find such effects.
Lastly, we note that recent experimental investigations \cite{Sarma98} of other transition metal oxides
have seen {\em precisely} this form of the density of states, on the metallic side, as the metal-to-insulator transition
is approached.

\begin{figure}[t]
\begin{center}
\includegraphics[width=10cm]{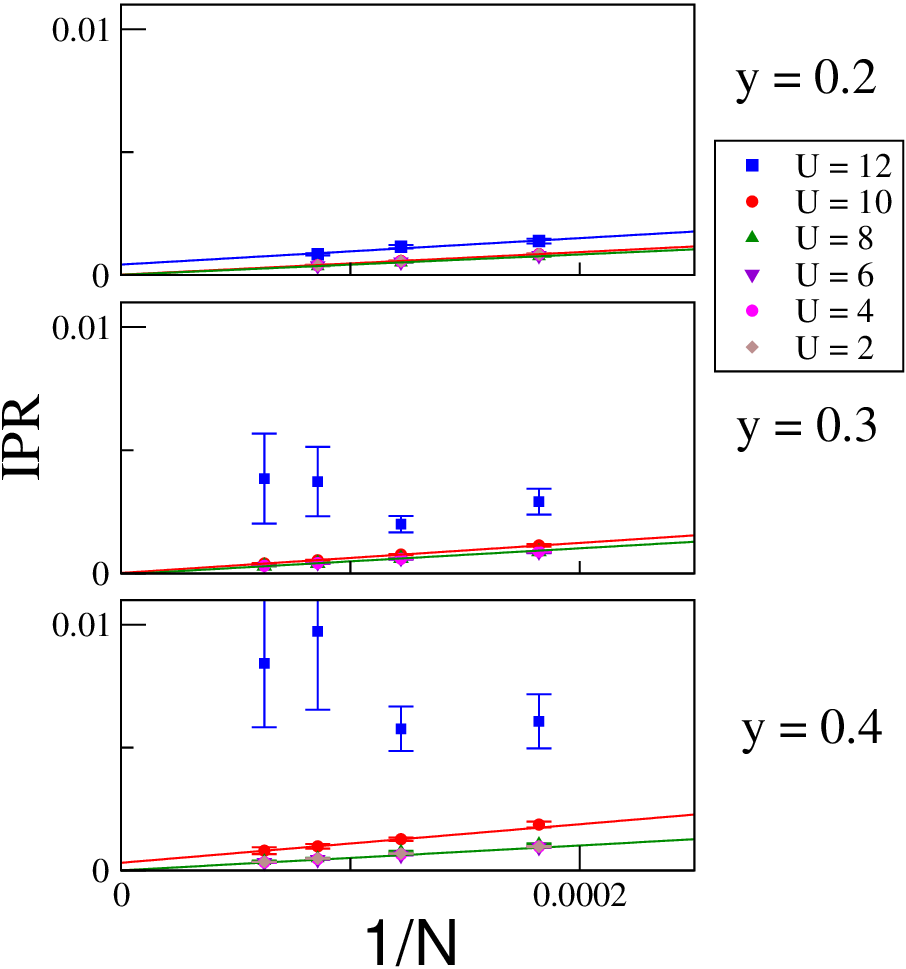}
\caption{\label{fig:IPR_X0pt234}Plots of the IPRs $vs.$ the inverse of the system size for the Hubbard+QSP model 
for the \LiAlTiO~ system with $y=0.2,0.3$ and 0.4 for different strengths of the Hubbard on-site interaction.
Note that the vertical scale in all of these plots is identical.}
\end{center}
\end{figure}

In order to characterize the metallic or insulating behaviour of such systems, we examined the 
eigenstates of a small energy bin ($\Delta E/t=0.1$) located symmetrically about the Fermi energy,
and calculated the average inverse participation ratio (IPR), averaged over a sufficient number of realizations
to obtain converged data. A collection of our results for the IPRs $vs.$ the inverse of the system size is
shown in Fig.~\ref{fig:IPR_X0pt234}. From the linear fits of the IPRs $vs.$ the inverse of the
system size, the values of the IPRs in the thermodynamic limit can be
extracted. We also determined the $98\%$ confidence level ($2 \times$ the standard deviation) error bars of
the intercepts.  As is well known, if the extrapolated value of the IPR has a finite intercept then the eigenstates 
corresponding to that specific system, in that energy range, are localized, and in the subsequent discussion
we use the criterion that the intercept must be at least twice the standard deviation above zero before we
classify (with a $98\%$ confidence level) that those energy eigenstates are of a localized nature. For example,
for $y=0.3$ for $U/t=2$ to $10$ the states near the Fermi level are extended, whereas for $U/t=12$ these
same states are localized. Consequently, as a function of increasing Hubbard interaction we identify the MIT
as occurring at $U_c/t \sim 11-12$ for $y=0.3$.

Repeating this sequence of calculations for the above-mentioned $y$ and $U/t$ we identified a
phase diagram of the ($T=0$) metal-to-insulator transition of this system with respect to disorder and interaction,
and our results are depicted in Fig.~3. We immediately see that if strong
electronic correlations are indeed an important aspect of the physics associated with this transition,
that is for a $y_c^{expt} \sim 0.33$, we require a $U/t\sim11-12$. Noting that for the noninteracting electrons
on this lattice one has a bandwidth of 8$t$, this implies that \LTO~ is a moderately correlated 
three-dimensional electronic system. This is consistent with conclusions drawn, albeit indirectly,
from recent low-temperature specific heat measurements\cite{Sun04} in this material.

In fact, this value is not far from experimental estimates of $U/t$ for the $t_{2g}$ $d$ orbitals of related $Ti$-based 
transition metal oxides. That is, 
the estimate for the Hubbard on-site repulsion for the $t_{2g}$ orbitals of Ti atoms in the perovskite LaTiO$_3$ is 
$U_{t_{2g}} \sim 3.1 eV$ \cite{Solovyev96}. Resonant soft-x-ray emission spectroscopy on perovskite-type Ti 
compound La$_x$Sr$_{1-x}$TiO$_3$ provides an estimate of $U_{d-d} \sim 4.0-4.4 eV$ \cite{Higuchi99,Higuchi03}, 
with the most recent estimate being $U/t \sim 4.0 eV$(2003)\cite{Higuchi03}. Now compare these estimates 
to our predicted value of $U$ for the \LiAlTiO~ system: taking into account the estimate for the transfer integral 
of $t \sim 0.33 eV$ for the $t_{2g}$ band in this system based on the LDA calculations\cite{Massidda88,Satpathy87},
using $U_c/t=11-12$ (for $y=0.3$) our estimate of $U$ corresponds to 3.7-4.0 eV.  Clearly, our estimate
is not dissimilar to the experimentally observed values for these related Ti-based oxides.

\begin{figure}[t]
\begin{center}
\includegraphics[width=8.5cm]{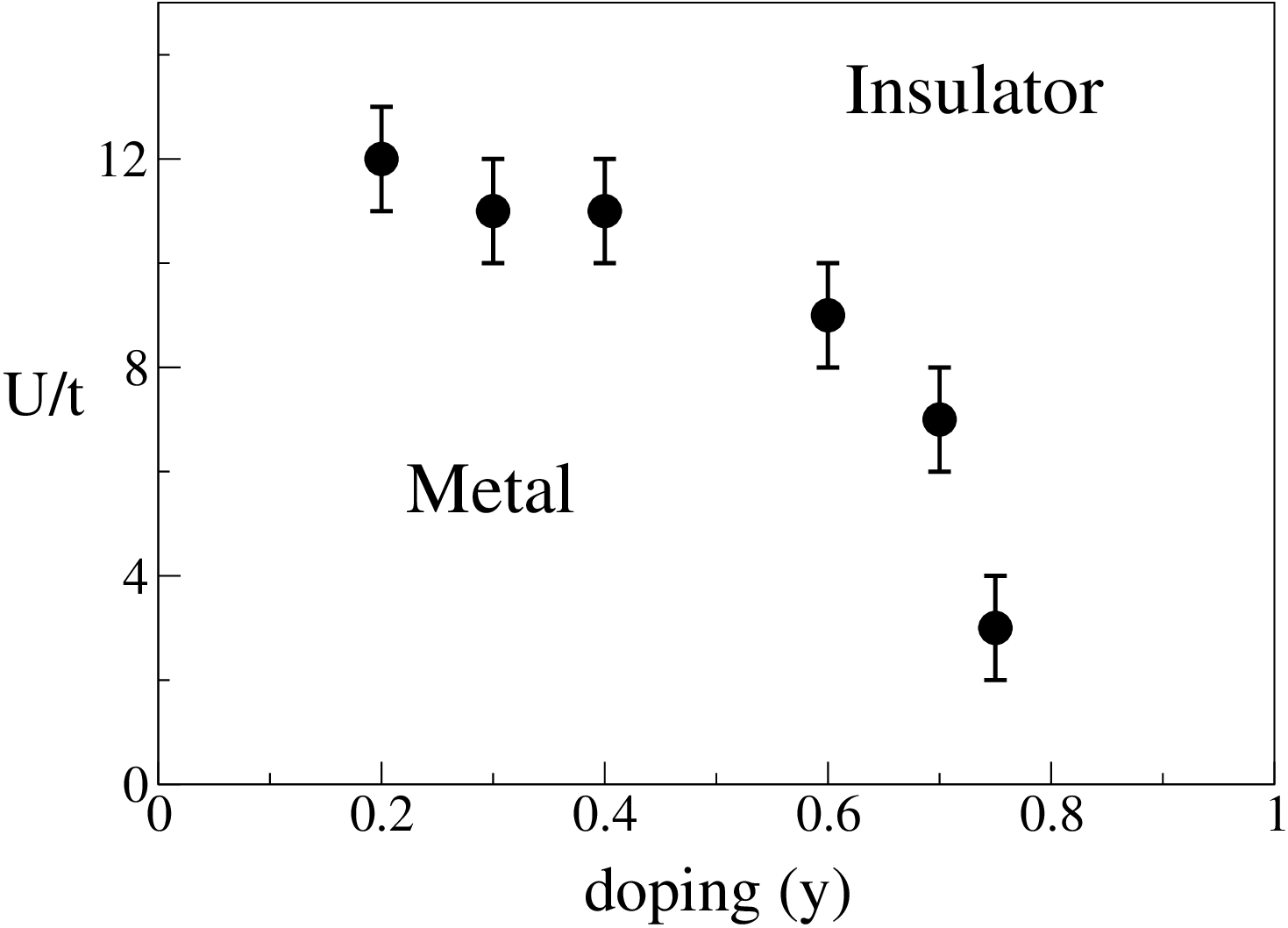}
\caption{\label{fig:MIT_phase_diagram}Phase diagram for the metal-insulator transition in the Hubbard+QSP model 
representative of the \LiAlTiO~system. Filled circles are estimates from the thermodynamic limit
extrapolation of the IPRs, discussed in the text.}
\end{center}
\end{figure}

We note that this energy leads to a provocative comparison between the near-neighbour superexchange ($J$) 
between moments in this system \textit{vs.} those in the high-temperature superconducting cuprates. That is,
using $J=4t^2/U$ for this one-band system we find that in \LiTiO~$J/t\approx 1/3$, which is similar
to the estimates for the cuprates. So, this similarity also lends support to the hypothesis of
Bednorz and M\"uller \cite{Muller96} of the potential relation of the pairing in this
system to the cuprates.

As further corroboration of our phase diagram results, 
and to gain a better understanding of how the transition is connected to the electronic 
and magnetic properties of this system, we have examined the (spin-resolved) charge and magnetic densities as a function
of disorder and Hubbard energy. Our results for the variation of the number of electrons per site are effectively
independent of spin, and are shown in Fig.~4. We see that for $y=0.3$, for $U/t = 0$ to 8, 
essentially no change takes place. However, as $U/t$ is increased from 8 to 10 to 12 this quantity undergoes 
substantial change. That is, when the Hubbard energy approaches the critical value (for this $y$) due to the 
proliferation of localized states arising from the suppression of the density of states at the Fermi level, 
one finds a far more inhomogeneous system. Also, our numerical results for the local magnetizations correspond to 
the appearance of (short-ranged) antiferromagnetic correlations {\em at} the metal-to-insulator transition. That is,
as $U/t$ is increased towards $U_c/t(y)$ this distribution gradually changes from a rather narrow peak at zero (paramagnetism), 
to a distribution dominated by two broad peaks at $\pm m_0$, with $m_0 \sim 0.4$,
corresponding to a locally antiferromagnetic arrangement \textit{but} with many sites having 
largely paramagnetic character.

\begin{figure}[t]
\begin{center}
\raisebox{0cm}[6.5cm][-.5cm]{
\includegraphics[width=10.0cm,height=7.cm]{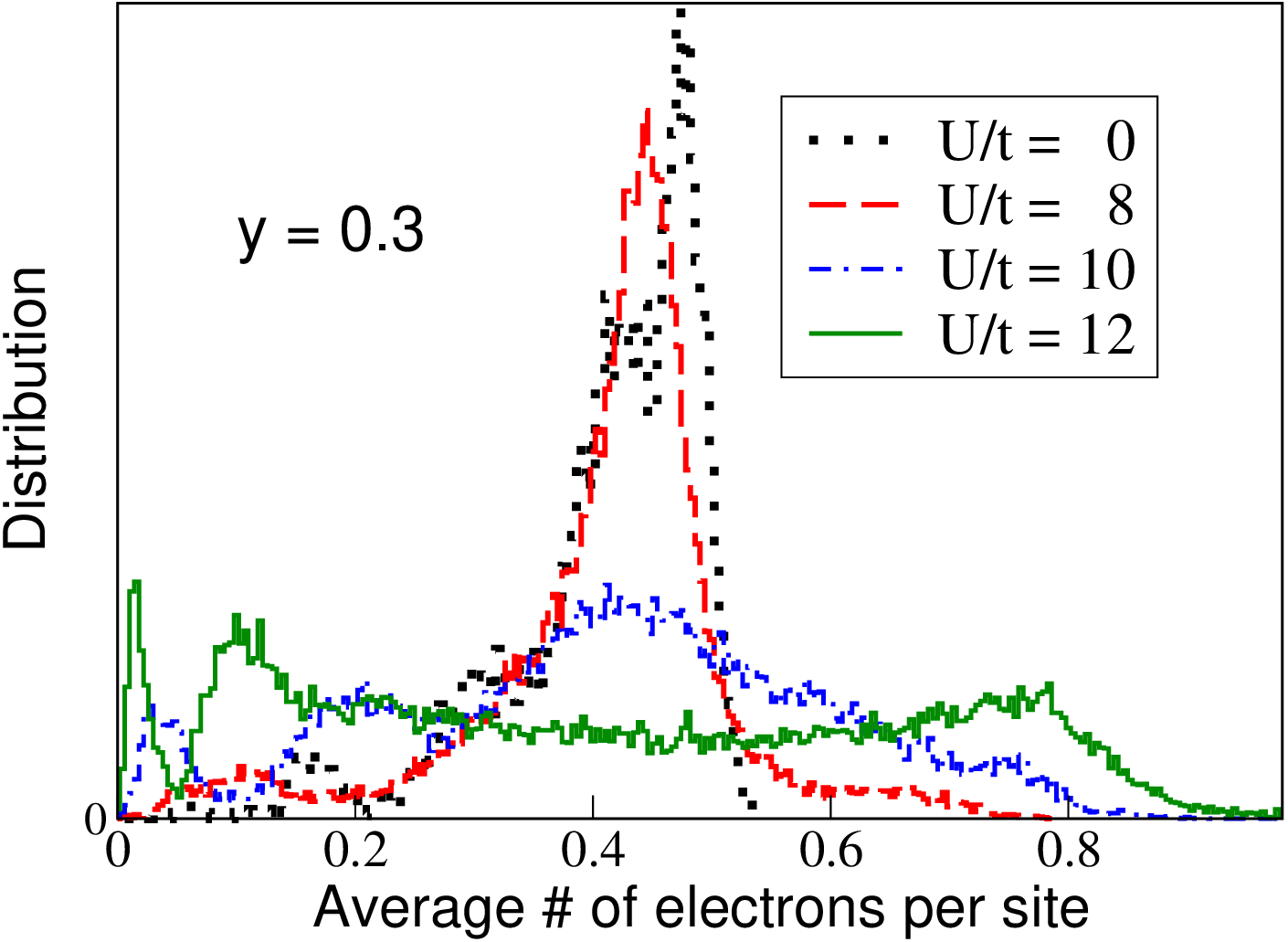}}
\caption{\label{fig:ni_withU} The distribution of single-site charge densities for $y=0.3$
as a function of increasing Hubbard energy $U/t$. One sees a profound change in
the homogeneity of the charge density as $U/t$ approaches the critical value at
which the metal-to-insulator transition occurs.}
\end{center}
\end{figure}

Summarizing, we have used a real-space self-consistent Hartree-Fock formulation to examine
the metal-to-insulator transition in disordered \LiAlTiO; this work (i) treats the disorder
exactly (that is, diagonalizing the full Hamiltonian matrix for any particular complexion of
disorder), and (ii) involves solving for the self-consistent solutions at every site. We have
found that to obtain agreement with the experimentally observed concentration of
Al impurities at which the metal-to-insulator transition occurs, a Hubbard energy 
somewhat larger than the noninteracting bandwidth is required, consistent with recent
experiments on this system\cite{Sun04}, and that the resulting 
density of states as the transition is approached is similar to that found in experiments 
on related materials \cite{Sarma98}.

We thank Nandini Trivedi and Barry Wells for helpful comments. This work was partially supported by NSERC of Canada and OGSST.
The work at Ames was supported by the USDOE under Contract No. W-7405-Eng-82.


\begin{thebibliography}{10}

\bibitem{Bednorz86}
J.~G. Bednorz and K.~A. M{\"u}ller,  Cond. Mat.: Z. Phys. B {\bf 64}, 189 (1986).

\bibitem{CMR-Dagotto}
E.~Dagotto, Science {\bf 309}, 257 (2005).

\bibitem{Moshopoulou99}
E.~G. Moshopoulou.
J. Am. Ceram. Soc. {\bf 82}, 3317 (1999).

\bibitem{Satpathy87}
S.~Satpathy and R.~M. Martin, Phys. Rev. B {\bf 36}, R7269 (1987).

\bibitem{Massidda88}
S.~Massidda, J.~Yu, and A.~J. Freeman,
Phys. Rev. B {\bf 38}, 11352 (1988).

\bibitem{Muller96}
K.~A. M{\"u}ller.
\newblock In B.~Batlogg et~al., editors, {\em Proceedings of the 10th
  Anniversary HTS Workshop on Physics, Materials, and Applications}, page~1.
  World Scientific, 1996.

\bibitem{Johnston76I}
D.~C. Johnston, J. Low Temp. Phys. {\bf 25}, 145 (1976).

\bibitem{Lambert90}
P.~M. Lambert, P.~P. Edwards, and M.~R. Harrison.
J. Sol. St. Chem. {\bf 89}, 345 (1990).

\bibitem{Kondo97}
S.~Kondo, D.~C. Johnston, et~al.,  Phys. Rev. Lett. {\bf 78}, 3729 (1997).

\bibitem{Fazileh04}
F.~Fazileh, R.~J. Gooding, and D.~C. Johnston,
Phys. Rev. B,{\bf 69}, 104503 (2004).

\bibitem{ACSTLcomment} The effects of the impurity-generated distribution of on-site energies are known to be very small -- F. Fazileh, X. Chen, R. J. Gooding and K. Tabunshchyk, Phys. Rev. B {\bf 73}, 035124 (2006).

\bibitem{Trivedi04}
  D. Heidarian and N. Trivedi, Phys. Rev. Lett. {\bf 93}, 126401 (2004).

\bibitem{SPINcomment} We note that this formulation is not invariant in spin space, an approximation that is acceptable for the low density systems 
(less than 1/4 filling) that we examine in this letter. 
We will present a detailed study of the role of
such terms in physical systems that are close to half filling in a future
publication -- X. Chen and R.~J. Gooding (unpublished).

\bibitem{Altshuler79}
B.~L.~Altshuler and A.~G.~Aronov,
Solid State Com. {\bf 30}, 115 (1979).

\bibitem{Sarma98}
D.~D.~Sarma, {\em et al.},
Phys. Rev. Lett. {\bf 80}, 4004 (1998).

\bibitem{Sun04}
C.~P.~Sun, {\em et al.}, Phys. Rev. B {\bf 70}, 054519 (2004).

\bibitem{Solovyev96}
I.~Solovyev, N.~Hamada, and K.~Terakura,
Phys. Rev. B {\bf 53}, 7158 (1996).

\bibitem{Higuchi99}
T.~Higuchi, {\em et al.},
Phys. Rev. B {\bf 60}, 7711 (1999).
                                                                                
\bibitem{Higuchi03}
T.~Higuchi, {\em et al.},
Phys. Rev. B {\bf 68}, 104420 (2003).

\end{thebibliography}
\end{document}